# Hidden antiferromagnetism on the light-irradiated surface of bulk SrTiO$_3$ and at the LaAlO$_3$/SrTiO$_3$ interfaces

Lev P. Gor'kov

*NHMFL, Florida State University, 1800 East Paul Dirac Drive, Tallahassee Florida 32310, USA and L.D. Landau Institute for Theoretical Physics of the RAS, Chernogolovka 142432, Russia*

We study properties of electrons on illuminated surfaces of SrTiO$_3$ with titanium $d_{xz}/d_{yz}$ and $d_{xy}$ bands for their spectrum. Recently A. F. Santander-Syro *et al* [Nature Materials, **13**, 1085 (2014)] found that the $d_{xy}$ bands actually comprise two chiral branches with the Kramers degeneracy at the zone center lifted *in absence of a magnetic moment*. From symmetry analysis of instabilities possible in the Fermi liquid with exchange interactions we identified the metallic in-layer state with the concrete *antiferromagnetic* phase and discuss if the same state materializes at conducting LaAlO$_3$/SrTiO$_3$ interfaces.

*Introduction.* The groundbreaking discovery by Ohtomo and Hwang [1] of a metallic electronic layer at interfaces between the two oxides LaAlO$_3$ (LAO) and SrTiO$_3$ (STO) lay off the new exciting field of oxide electronics. A tunable two-dimensional gas of electrons (2DEG) with the surface density of charge in the range $n_s \sim 10^{13} \div 10^{15} cm^2$ and a mobility as high as few $10^3 cm^2/V\sec$ presents a perfectly new conducting liquid thus opening high expectations for the technologic advance [2]. Further developments have lead to disclose at the LAO/STO interfaces of a variety of remarkable phenomena of great interest for the fundamental condensed matter physics in general, including, in many instances *coexisting*, 2D ferromagnetism [3-5] and 2D superconductivity [6-11].

A decade-long work using various advanced experimental techniques in the combination with theoretical calculations allowed to reconstruct structure of the quantum well at the intersection of such two wide-band insulators reasonably well. Yet, no major breakthrough at the level of microscopic physics was achieved regarding properties of the in-layer electronic gas. In particular, it concerns the nature of the insulator-to-metal transition. The latter bears the threshold character and occurs when thickness $d$ of the LaAlO$_3$ layer on top of SrTiO$_3$ surpasses the value $d = d_{cr}$ of 4unit cells (u.c.) [12].

The situation is changing with discovery of a metallic layer on the illuminatedTiO$_2$-terminated surface of bare SrTiO$_3$; pure crystalline SrTiO$_3$ being insulator with a broad gap $\approx 3.5 eV$, electrons are doped into the bending conduction band irradiating surface by the ultraviolet light [13-15]. The spectrum of 2DEG on the irradiated surface is comprised of the two groups of bands formed of the titanium *3d*-levels: the two light $d_{xy}$ bands and pair of the heavier $d_{xz}/d_{yz}$ ones. The four Fermi surfaces seen by ARPES are two concentric rings originating from the $d_{xy}$-bands and two $d_{xz}/d_{yz}$-ellipsoids elongated along the $k_x$-and $k_y$-directions [14, 15].

Recent *spin and angle* resolved photoemission (SARPES) experiments[16] revealed that these two light $d_{xy}$-bands in reality are a pair of the *spin-polarized* branches with spins winding in the opposite directions, as in case of a surface band in the presence of the Rashba spin-orbit interaction[17]. Surprisingly, the two branches *seem to be split* at the 2D Brillouin zone center with the Kramers degeneracy at the momentum $p=0$ lifted. That is, the time reversal invariance is broken without a visible magnetic moment.

The broken time-reversal symmetry in the $d_{xy}$ band [16] signifying a hidden magnetism is most intriguing. Below we propose the interpretation in terms of the Fermi liquid effects. We argue that the exchange interactions between electrons in the surface layer make the latter unstable towards spontaneous transition into a magnetic phase. Gaining in energy, in the new phase 2DEG acquires stability to disorder related to oxygen vacancies and other irregularities at the surface. Our symmetry analysis suggests the *antiferromagnetic* order parameter in form of a vector perpendicular to the plane and depending on the azimuth angle. Consistent with anisotropy of the out-of-plane component of the spins polarization, such order parameter leads to minor changes in shape of the two concentric $d_{xy}$–Fermi surfaces. We show that our results are compatible with the accuracy of data [16, 18].

The outstanding question then is whether the metallic energy spectrum observed on the *irradiated* surface of $SrTiO_3$ may have any relation to the spectrum of 2DEG in hetero-structures. Main experimental facts do not contradict the idea that electrons supplied by overlaying LAO layers onto *the Ti 3d-levels* of $SrTiO_3$ form at metal-to-insulator transition 2DEG same as on illuminated surfaces of *bare* $SrTiO_3$ keeping the topology of the $d_{xy}$-and $d_{xz}/d_{yz}$-Fermi surfaces intact.

*Evidences of electron-electron interactions.* Strong exchange correlations between electrons seem playing a role in stability of 2DEG on the illuminated $SrTiO_3$ [16]. Large band splitting $2S \approx 90 meV$ corroborates idea. Electron-electron correlations were proven to contribute at the analysis of the STM data [19].

Most explicitly, the Fermi liquid effects show up in non-vanishing spectral weight seen by ARPES at higher binding energies [14]. With interactions the coherent quasi-particles spectrum below the Fermi surface gives way to spectrum of incoherent excitations and indeed, in Fig.1a [14] and Fig.1e, h [16] (at the Dirac point and below) ARPES spectra are significantly smeared by incoherent contributions.

Besides, the very existence of a highly conducting metallic layer in spite of inevitable disorder caused by generated oxygen vacancies [14, 15] speaks in favor that interactions between electrons prevail over the tendency to weak localization.

*Energy spectrum of 2D magnetic phases.* Following [16], consider the Hamiltonian:

$$\hat{H} = \vec{p}^2/2m^* + \alpha(p_x\hat{\sigma}_y - p_y\hat{\sigma}_x) + \hat{\sigma}_z S. \quad (1)$$

Here $m^*$ is the effective mass for a parabolic band; the vector $\vec{p}$ lies in the plane. The second term is the Rashba spin-orbit interaction [17]; $\hat{\sigma}_{x,y,z}$ are the Pauli spin matrices, $S$ is the perpendicular-to- plane component of some Zeeman field. (Dimensionless parameter $\kappa = m^*\alpha/p_F$ is a measure of strength of the Rashba term in Eq. (1)). The spectrum consists of the two branches $\lambda = (\pm)$:

$$E_\lambda(p) = \vec{p}^2/2m^* + \lambda\sqrt{\alpha^2\vec{p}^2 + S^2}. \quad (2)$$

As in [16], one finds that chirality has the opposite signs on two ($\lambda = \pm$) branches. For the component of electronic spin $<s_z>$ it follows:

$$<s_z> = \lambda S/2\sqrt{\alpha^2\vec{p}^2 + S^2}. \quad (3)$$

Beside that ferromagnetism has never been confirmed in $SrTiO_3$, non-zero magnetic moment would inevitably lead to formation of ferromagnetic domains; the local contributions from the non-zero moments are then expected to be averaged to zero in SARPES experiments [16].

According to [16], Eq. (2) reproduces the experimental spectrum reasonable well. We find out that Eq. (3) is in contradiction with certain experimental result significance of which, as it seems, went unnoticed there.

In fact, in Fig.3 c-f[16] the $z$-component, $<s_z>$ has different sign *for each* of the two pairs (*1,4*) and (*2,3*) shown in Fig.3 a[16], i.e., on the opposite side *of one and the same* branch; in Eq. (3) the sign of $<s_z>$ depends only on choice of the branch. Such behavior as in Fig.3 [16] is not compatible with a constant Zeeman vector $S$ in Eq. (1).

Below we propose the spectrum for the $d_{xy}$-band also in form of Eq. (2), but in which a constant vector $S$ is substituted with the vector $\bar{S}_E(\vec{p})$ depending on the azimuth angle $\varphi$ as:

$$\bar{S}_E(\vec{p}) = |S|\sin\varphi. \quad (4)$$

*Stability of 2DEG.* Phase transition that may occur in the electronic Fermi liquid can be listed invoking the so-called Pomeranchuck instabilities [19, 20]. We search for order parameters violating the time-reversal invariance; only phases with the *unchanged* lattice periodicity are considered below.

In the Hamiltonian for exchange interactions in the system of the form:

$$\hat{H}_{exc} = \sum_{\alpha\beta\lambda\delta} \iiint I(\vec{p},\vec{p}')([\hat{a}_\alpha^+(\vec{p})\vec{\hat{\sigma}}_{\alpha\beta}\hat{a}_\beta(\vec{p}-\vec{k})]\cdot[\hat{a}_\gamma^+(\vec{p}')\vec{\hat{\sigma}}_{\lambda\delta}\hat{a}_\delta(\vec{p}'+\vec{k})])d^2p\,d^2p'd^2k \quad (5)$$

assume a certain crystalline symmetry for the interaction $I(\vec{p},\vec{p}')$ and expand it at the Fermi surface over the normalized functions $\chi^{l,t}(\vec{p})$ belonging to all irreducible representations (*l*) of the given group (the group $C_{4v}$ in our case):

$$I(\vec{p},\vec{p}') = \sum_{t,l} I_l \chi^{l,t}(\vec{p})\chi^{l,t}(\vec{p}'). \quad (6)$$

(*t* numerates all basis functions if a representation were degenerate). The general form for the magnetic order parameter is a certain vector $\vec{S}_l(\vec{p})$:

$$\vec{S}_l^i(\vec{p}) = \sum_t \eta_t^{l,i} \chi^{l,t}(\vec{p}). \quad (7)$$

(Vectors $\vec{S}_l(\vec{p})$ in (7) originate from one of the "anomalous" averages of the particle operators in Eq. (5) $\hat{a}_\alpha^+(\vec{p})\hat{\sigma}_{\alpha\beta}^i\hat{a}_\beta(\vec{p}) \Rightarrow \int_0^{p_F} <\hat{a}_\alpha^+(\vec{p})\hat{\sigma}_{\alpha\beta}^i\hat{a}_\beta(\vec{p})>(m^*dp/2\pi)$ integrated over $|\vec{p}|$ and multiplied by one of the interaction constants $I_l$ in (6); index (*i*) numerates components of the vector).

We consider only vectors $S_z(\vec{p})$ *perpendicular*- to-plane (the in-plane vector $\vec{S}$ cannot split the Dirac point) and a *non-identical* representation $\chi^{l,t}(\vec{p})$. If averaged over the in-plane angle, $<S_z(\vec{p})> \Rightarrow \int \chi^{l,t}(\vec{p})d\varphi \equiv 0$; as distinct from the ferromagnetic vector (at the identical representation), such "*itinerant antiferromagnetic*" order parameter does not imply the domains' formation.

For the derivation below, recall that in the Landau Fermi liquid theory energy $\varepsilon_{\alpha\beta}(\vec{p})$ of an elementary excitation is made up of a "bare" energy $\varepsilon_{0,\alpha\beta}(\vec{p})$ and of the second contribution accounting for action on the part of all particles *disturbed by emergence* of the excitation $\varepsilon_{\alpha\beta}(\vec{p})$:

$$\varepsilon_{\alpha\beta}(\vec{p}) = \varepsilon_{0,\alpha\beta}(\vec{p}) + \int f_{\alpha\beta;\gamma\delta}(\vec{p};\vec{p}')[\delta n(p')]_{\delta\gamma} d^2p'/(2\pi)^2 \quad (8)$$

In the notations $f(\vec{p};\vec{p}'|\sigma;\sigma') = \varphi(\vec{p};\vec{p}') + (\vec{\sigma}\cdot\vec{\sigma}')\xi(\vec{p};\vec{p}')$ the second term stands for spin-spin interactions in the *exchange* approximation; $\xi(\vec{p};\vec{p}')$ can be expanded over same irreducible representations as in Eq. (6):

$$\xi(\vec{p},\vec{p}') = \sum_{t,l} Z_l \chi^{l,t}(\vec{p})\chi^{l,t}(\vec{p}'), \quad (9)$$

To *probe stability* of the system, let $\hat{\sigma}_z S_l(\vec{p})$ a small term having the form of Eq. (7) is added to a "bare" $\varepsilon_{0,\alpha\beta}(\vec{p})$. Such perturbation causes in return a change $\hat{\varepsilon}(\vec{p}) \Rightarrow \hat{\varepsilon}(\vec{p}) + \hat{\sigma}_z \bar{S}_l(\vec{p})$ in energy of the excitation $\varepsilon_{\alpha\beta}(\vec{p})$. The relation between $S_l(\vec{p})$ and $\bar{S}_l(\vec{p})$ follows from (8):

$$\hat{\sigma}_z \bar{S}_l(\vec{p}) = \hat{\sigma}_z S_l(\vec{p}) + \int (\vec{\sigma} \cdot \vec{\sigma}'_{\gamma\delta}) \xi(\vec{p}; \vec{p}') [\delta n(p')]_{\delta\gamma} d^2 p' / (2\pi)^2. \quad (10)$$

(Here $[\delta n(p')]_{\delta\gamma}$ means the difference $[\delta n(p')]_{\delta\gamma} = n[\varepsilon_0(\vec{p}) + \sigma_z \bar{S}_l(\vec{p})]_{\delta\gamma} - n[\varepsilon_0(\vec{p})]$; $n[\varepsilon_0(\vec{p})]$ is the Fermi function). In the right hand side of Eq. (10) $[\delta n(p')]_{\delta\gamma}$ can be expanded in powers of $\bar{S}_l(\vec{p})$. Leaving at the moment only the linear in $\bar{S}_l(\vec{p})$ term:

$$\bar{S}_l(\vec{p}) = S_l(\vec{p}) + \bar{S}_l(\vec{p}) Z_l \sum_\lambda \int (\partial n_0(p') / d\varepsilon)(p' dp' / \pi) \quad (11)$$

one finds:

$$\bar{S}_l(\vec{p}) = S_l(\vec{p}) / \left[1 + Z_l \sum_\lambda \nu_\lambda(\varepsilon_F)\right] \equiv S_l(\vec{p}) / (1 + \bar{Z}_l). \quad (12)$$

In (12) $\bar{Z}_l$ is $Z_l$ multiplied by the summary density of states $\nu(\varepsilon_F)$. (On each Fermi surface $pdp/\pi \Rightarrow (p/\pi)(dp/d\varepsilon)d\varepsilon = (pm^*/\pi(p + \lambda\alpha m^*))d\varepsilon \equiv \nu_\lambda(\varepsilon)d\varepsilon$; $\lambda = \pm$ stands for the two solutions $\varepsilon_\pm(p) = \vec{p}^2 / 2m^* \pm \alpha p$; $p_{F,\lambda}$ are the two Fermi momenta).

The pole at $\bar{Z}_l = -1$ would signify instability of the system with respect to the spontaneous transition into a phase with one of vectors $\vec{S}_l(\vec{p})$ in (6) as the order parameter.

Vector $\bar{S}_l(\vec{p})$ substituted, Eqs. (2, 3) acquire dependence on the azimuth angle. Besides the identical representation $x^2 + y^2$, the group $C_{4v}$ has two one-dimensional representations: $B_1: xy \propto \sin 2\varphi$ and $B_2: x^2 - y^2 \propto \cos 2\varphi$, and one two-dimensional representation: $E: x, y \Rightarrow \cos\varphi, \sin\varphi$. $B_1$ and $B_2$ are both even at $\varphi \Rightarrow \varphi \pm \pi$. Fig.3 c-f [16] for $<s_z>$ seems to be consistent only with choice of the $E$-representation. Of the two components ($\cos\varphi, \sin\varphi$) one must choose one; in correspondence with Fig.3 c-f[16] we suggested $\bar{S}_E(\vec{p}) = |S| \sin\varphi$ in Eq.(4).

Substitution of (4) into (2) gives for the energy spectrum:

$$E_\lambda(p) = \vec{p}^2 / 2m^* + \lambda \sqrt{\alpha^2 \vec{p}^2 + S^2 \sin^2 \varphi}. \quad (13)$$

Near the zone center: $E_\lambda(p) \approx \lambda\sqrt{\alpha^2 \vec{p}^2 + S^2 \sin^2\varphi}$. Let $E > 0$. Projections of the contours of constant energy $E$ in Fig.1a encircle figures oblate along the $y$-axis. A 3D view of the spectrum $E_+(\vec{p})$ near the zone center is shown in Fig. 1b.

*Shape of the $d_{xy}$-Fermi surfaces.* Eq. (13) implies the two-fold axial symmetry also for the Fermi surfaces of the two $d_{xy}$-bands. To estimate the magnitude of deviations from ring-like shape of two concentric Fermi surfaces in the isotropic spectrum Eq. (2) with $S = const$, rewrite the difference $\vec{p}_{F,\lambda}^2(\varphi) - p_F^2 = \lambda\sqrt{(2m^*\alpha p_{F,\lambda})^2 + (2m^*S)^2 \sin^2\varphi}$ in the following dimensionless form: $\vec{p}_{F,\lambda}^2(\varphi)/p_F^2 - 1 = \lambda\sqrt{(2\alpha m^*/p_F)^2 + (S/\varepsilon_F)^2 \sin^2\varphi}$; $p_{F,\lambda}(\varphi)$ are the Fermi momenta on the two branches $\lambda = (\pm)$; difference between $p_{F,\lambda}(\varphi)$ and $p_F$ under the square root is neglected. Taking for purpose of illustration all parameters $m^* = 0.65 m_e$, $\alpha = 5 \cdot 10^{-11} eVm$, $k_{F,+} \simeq 0.18 \cdot A^{-1}$, $k_{F,-} \simeq 0.12 \cdot A^{-1}$; $k_F = (k_{F,+} + k_{F,-})/2 \simeq 0.15 \cdot A^{-1}$, and $S \sim 40 meV$ from [16] and substituting numbers, we find $|k_{F,\lambda}(\pi/2) - k_{F,\lambda}(0)|/2 \approx 0.05 k_F \sim 0.0075 A^{-1}$. That is, deviations from shape of a ring are small and choice of the order parameter Eq. (4) would not contradict experimental data [16].

*Mean field theory near a threshold* ($|\bar{Z}_{0,l} + 1| \ll 1$). It is instructive to consider the transition near threshold of the phase stability. To proceed, omit "bare" $S_l(\vec{p})$ in Eq. (11) and expand $[\delta n(p')]_{\delta\gamma}$ up to terms of the third order in $\bar{S}_l(\vec{p})$; multiply both sides in turns by functions $\chi^{l,t}(\vec{p})$ and integrate along the Fermi surface. Such common procedure leads to equations for the set of parameters $\eta_t$ in Eq. (6). For the two-dimensional representation $E: x, y \Rightarrow \cos\varphi, \sin\varphi$ the general form of $\bar{S}_E(\vec{p})$ is $\bar{S}_E(\vec{p}) = \eta_1 \cos\varphi + \eta_2 \sin\varphi$. To determine correspondence between parameters $(\eta_1, \eta_2)$ one needs, as in the Landau theory of the second order phase transitions, a functional $F(\eta_1, \eta_2)$ to present the two equations for $(\eta_1, \eta_2)$ in the equivalent form as the equations of equilibrium $\delta F(\eta_1; \eta_2)/\delta\eta_t = 0$. Simple but tedious calculations give $F(\eta_1, \eta_2) \propto \{(1 + \bar{Z}_E)(\eta_1^2 + \eta_2^2) + (A/4\varepsilon_F^2)[(\eta_1^2 + \eta_2^2)^2 - (1/2)(\eta_1^4 + \eta_2^4)]\}$. (For definition of A, see below). Minimum in $F(\eta_1, \eta_2)$ (if $A > 0$) is reached at the non-zero either $(\eta_1)$, or $(\eta_2)$. In conjunction with data in Fig.3 c-f [16] we choose $\eta_1 = 0$. With the term quadratic in temperature one obtains:

$$|1 + \bar{Z}_E| = (A/8)[(\bar{S}_E^2 + T^2)/\varepsilon_F^2]. \qquad (14)$$

In (14) $A = -\Sigma_2/\nu(\varepsilon_F)$, $\Sigma_2 = \varepsilon_F^2 \sum_\pm (\partial^2 \nu_i(\varepsilon)/\partial\varepsilon^2)$ sums the second derivatives of density of states on two branches, $\varepsilon_F = p_F^2/2m^*$).

In three dimensions $\partial^2 \nu(\varepsilon)/\partial\varepsilon^2 = -(1/4\varepsilon^2)$ and one may infer the second order character of the phase transition at the Pomeranchuk instability. The 2D density of states is a constant; the second derivative equals zero identically even for the energy spectrum with the Rashba term $E_\lambda(p) = \vec{p}^2/2m^* + \lambda\alpha p$. A small cubic term if added into the spin-orbit interaction $\alpha p \Rightarrow \alpha_0 p(1+\gamma p^2/p_F^2)$ produces a non-zero $\Sigma_2 \propto -\alpha_0\gamma$. The sign is not theoretically determined. At $\Sigma_2 < 0$ Eq. (14) describes the second order phase transition. At $\Sigma_2 > 0$ the transition would be of the 1$^{st}$ order and higher terms in the expansion become necessary.

The *only characteristic energy scale* at a phase transition driven by the Pomeranchuk instability mechanism is the Fermi energy $\varepsilon_F = p_F^2/2m^*$. One infers that in the magnetic phase $2\bar{S}_E \sim T_c \sim \varepsilon_F \sim 0.1 eV$, as indeed found in [16].

*Discussion of nature of 2DEG at the LAO/STO interface*. Metal-insulator transition at LAO/STO interfaces bears the abrupt character. Thus, at the 4 unit cell-thick LaAlO$_3$ layer the conductance jumps by five orders in the magnitude from below the measurable limit [12]. Non-zero density of states is first seen by STM at same thickness [21]. Four unit cells (u.c.) of LAO are critical both for ferromagnetism [22, 23] and for 2D superconductivity [24, 25]). The metal-insulator transition seems to coincide with a *structural change* occurring *at 4 u.c.* described in [18] as a polar Ti-O-buckling. Such sharp a threshold reminds of the 1$^{st}$ order-type phase transition.

By the spectroscopic means 2DEG on the *buried* LAO /STO interfaces was studied in [11, 18]. Mapped in the *k*-space [11], the band states at the interface with LAO layers exceeding the critical thickness 4 u.c. reveal same topology of the Fermi surfaces as in [13]. Another instructive fact is this. As process of doping in the LAO /STO interfaces generally goes with increasing thickness of the LAO layer, both the itinerant and localized electrons in all cases occupy the same *Ti 3d-levels on the side of SrTiO$_3$* [11, 26- 27]. Taken together, the above experimental facts give strong support to the idea that electrons at the irradiated SrTiO$_3$ surface and at the LAO/STO interfaces undergo kind of the 1$^{st}$ order transformation into *one and the same phase* of two-dimensional electronic Fermi liquid with the reduced magnetic symmetry.

*In summary*, magnetism of the conducting layer was inferred from the fact of lifted Kramers degeneracy in [16]. We proposed the Pomeranchuck-type instability as the mechanism for formation of the magnetic phase and pointed out the concrete symmetry of the *antiferromagnetic* order parameter. We argued that a large energy scale of order of the Fermi energy inherent in this mechanism protects the ground state of 2DEG against ever-present random effects of disorder. The theoretical concepts put forward in the above prompt for further elaboration on the part of experiment. In particular, we suggest analysis of the perpendicular-to-plane spin polarization component. It would be equally interesting if ARPES were able to discern the two-fold symmetry in vicinity of the Dirac point at the center of the Brillouin zone.

**Acknowledgment** The author thanks L. Levitov for drawing his attention to experiments [16] and E. Rashba for many stimulating discussions and G. Teitel'baum for his help with Figure 1. The work was supported by the NHMFL through NSF Grant No. DMR-1157490, the State of Florida and the U.S. Department of Energy.

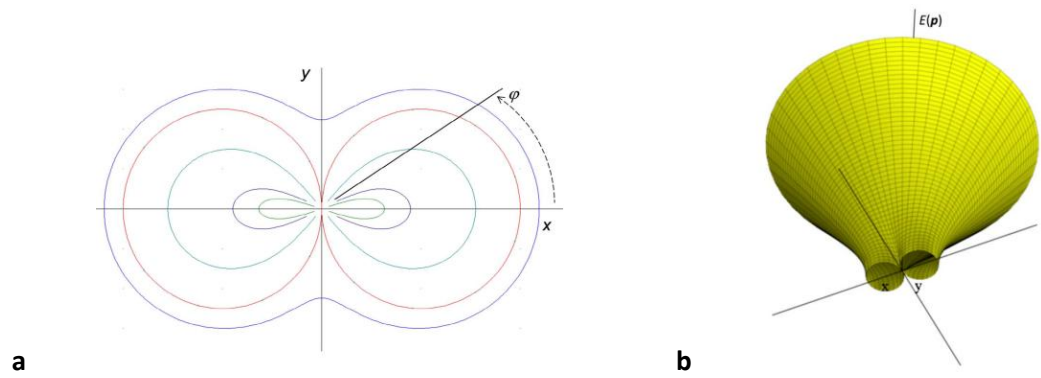

**Fig.1** The antiferromagnetic energy spectrum (13) $E_\lambda(p) \approx \sqrt{\alpha^2 \vec{p}^2 + S^2 \sin^2 \varphi}$ drawn in a close vicinity of the Brillouin zone center (for the branch with $E > 0$). **a**, Projections of contours of the constant-energy at different $E > 0$. **b,** A three-dimensional view (from the side of negative $E$).